\documentclass[onecolumn,3p]{elsarticle}
\usepackage{graphicx}
\usepackage{amssymb}
\usepackage{amsmath}

\journal{Computer Physics Communications}

\begin{document}

\begin{frontmatter}

\title{C programs for solving the time-dependent Gross-Pitaevskii 
equation in a fully anisotropic trap\tnoteref{funding}}
\tnotetext[funding]{D.V., I.V., and A.B.~acknowledge support by the 
Ministry of Education and Science of the Republic of Serbia under 
projects No.~ON171017 and NAD-BEC, by DAAD - German Academic and 
Exchange Service under project NAD-BEC, and by the European Commission 
under EU FP7 projects PRACE-1IP, PRACE-2IP, HP-SEE, and EGI-InSPIRE.
P.M. acknowledges support by DST and CSIR of India.
S.K.A. acknowledges support by FAPESP and CNPq of Brazil.}

\author[scl-ipb]{Du\v{s}an Vudragovi\'{c}}

\author[scl-ipb]{Ivana Vidanovi\'{c}}

\author[scl-ipb]{Antun Bala\v{z}\corref{cor1}}
\ead{antun@ipb.ac.rs}

\author[bdu]{Paulsamy Muruganandam}

\author[ift]{Sadhan K. Adhikari}

\address[scl-ipb]{Scientific Computing Laboratory, Institute of Physics Belgrade, University of Belgrade, Pregrevica 118, 11080 Belgrade, Serbia}
\address[bdu]{School of Physics, Bharathidasan University, Palkalaiperur Campus, Tiruchirappalli -- 620024, Tamil Nadu, India}
\address[ift]{Instituto de F\'{\i}sica Te\'{o}rica, UNESP -- Universidade Estadual Paulista,  01.140-70 S\~{a}o Paulo, S\~{a}o Paulo, Brazil}
\cortext[cor1]{Corresponding author}

\begin{abstract} 

We present C programming language versions of earlier published 
Fortran programs (Muruganandam and Adhikari (2009) [1]) for calculating both stationary and non-stationary 
solutions of the time-dependent Gross-Pitaevskii (GP) equation. The GP 
equation describes the properties of dilute Bose-Einstein condensates at 
ultra-cold temperatures. C versions of programs use the same algorithms 
as the Fortran ones, involving real- and imaginary-time propagation 
based on a split-step Crank-Nicolson method. In a one-space-variable 
form of the GP equation, we consider the one-dimensional, 
two-dimensional, circularly-symmetric, and the three-dimensional 
spherically-symmetric harmonic-oscillator traps. In the 
two-space-variable form, we consider the GP equation in two-dimensional 
anisotropic and three-dimensional axially-symmetric traps. The 
fully-anisotropic three-dimensional GP equation is also considered.
In addition to these twelve programs, 
for six algorithms that involve two and three space variables, we have also 
developed threaded (OpenMP parallelized) programs, which allow 
numerical simulations to use all available CPU cores on a computer. All
18 programs are optimized and accompanied 
by makefiles for several popular C compilers. We 
present typical results for scalability of threaded codes and 
demonstrate almost linear speedup obtained with the new programs, 
allowing a decrease in execution times by an 
order of magnitude on modern multi-core computers.

\end{abstract}


\end{frontmatter}

\begin{small}
\noindent
{\bf New version program summary}
\\

\noindent
{\em Program title:} GP-SCL package, consisting of: (i) imagtime1d, (ii) imagtime2d, (iii) imagtime2d-th, (iv) imagtimecir, (v) imagtime3d, (vi) imagtime3d-th,
(vii) imagtimeaxial, (viii) imagtimeaxial-th, (ix) imagtimesph, (x) realtime1d, (xi) realtime2d, (xii) realtime2d-th, (xiii) realtimecir, (xiv) realtime3d, (xv) realtime3d-th,
(xvi) realtimeaxial, (xvii) realtimeaxial-th, (xviii) realtimesph\\
{\em Journal Reference:} Comput. Phys. Commun. 183 (2012) 2021.    \\
{\em Catalogue identifier:} AEDU\_v2\_0       \\
{\em Program Summary URL:} http://cpc.cs.qub.ac.uk/summaries/AEDU\_v2\_0.html\\
{\em Program obtainable from:} CPC Program Library, Queen's University of Belfast, N. Ireland\\
{\em Licensing provisions:} Standard CPC licence, http://cpc.cs.qub.ac.uk/licence/licence.html\\
{\em No. of lines in distributed program, including test data, etc.:} 180~583\\
{\em No. of bytes in distributed program, including test data, etc.:} 1~188~688\\
{\em Distribution format:} tar.gz\\
{\em Programming language:} C and C/OpenMP\\
{\em Computer:} Any modern computer with C language compiler installed\\
{\em Operating system:} Linux, Unix, Mac OS, Windows\\
{\em RAM:} Memory used with the supplied input files: 2-4 MByte (i, iv, ix, x, xiii, xvi, xvii, xviii), 8 MByte (xi, xii), 32 MByte (vii, viii), 80 MByte (ii, iii), 700 MByte (xiv, xv), 1.2 GByte (v, vi)\\
{\em Number of processors used:} For threaded (OpenMP parallelized) programs, all available CPU cores on the computer.\\
{\em Classification:} 2.9, 4.3, 4.12\\
{\em Catalogue identifier of previous version:} AEDU\_v1\_0\\
{\em Journal reference of previous version:} Comput. Phys. Commun. 180 (2009) 1888.\\
{\em Does the new version supersede the previous version?:} No\\
{\em Nature of problem:} These programs are designed to solve the 
time-dependent Gross-Pitaevskii (GP) nonlinear partial differential equation 
in one-, two- or three-space dimensions with a harmonic, 
circularly-symmetric, spherically-symmetric, axially-symmetric or fully 
anisotropic trap. The GP equation describes the properties 
of a dilute trapped Bose-Einstein condensate.\\
{\em Solution method:} The time-dependent GP equation is 
solved by the split-step Crank-Nicolson method by discretizing in space 
and time. The discretized equation is then solved by propagation, in 
either imaginary or real time, over small time steps. The method yields 
solutions of stationary and/or non-stationary problems.\\
{\em Reasons for the new version:} Previous Fortran programs [1] are 
used within the ultra-cold atoms [2-11] and nonlinear optics [12,13] 
communities, as well as in various other fields [14-16]. This new 
version represents translation of all programs to the C programming 
language, which will make it accessible to the wider parts of the 
corresponding communities. It is well known that numerical simulations of the GP equation in 
highly experimentally relevant geometries with two or three space 
variables are computationally very demanding, which presents an obstacle 
in detailed numerical studies of such systems. For this reason, we have 
developed threaded (OpenMP parallelized)
versions of programs imagtime2d, imagtime3d, 
imagtimeaxial, realtime2d, realtime3d, realtimeaxial, which are named 
imagtime2d-th, imagtime3d-th, imagtimeaxial-th, realtime2d-th, 
realtime3d-th, realtimeaxial-th, respectively. Figure 1 shows the 
scalability results obtained for OpenMP versions of programs realtime2d 
and realtime3d. As we can see, the speedup is almost linear, and on a 
computer with the total of 8 CPU cores we observe in Fig.~1(a) a
maximal speedup of around 7, 
or roughly 90\% of the ideal speedup, while 
on a computer with 12 CPU cores we find in Fig.~1(b) that 
the maximal speedup is around 9.6, or 80\% of the ideal speedup.
Such a speedup represents significant improvement in the performance.\\
\begin{figure}[!ht]
\begin{center}
\includegraphics[width=7cm]{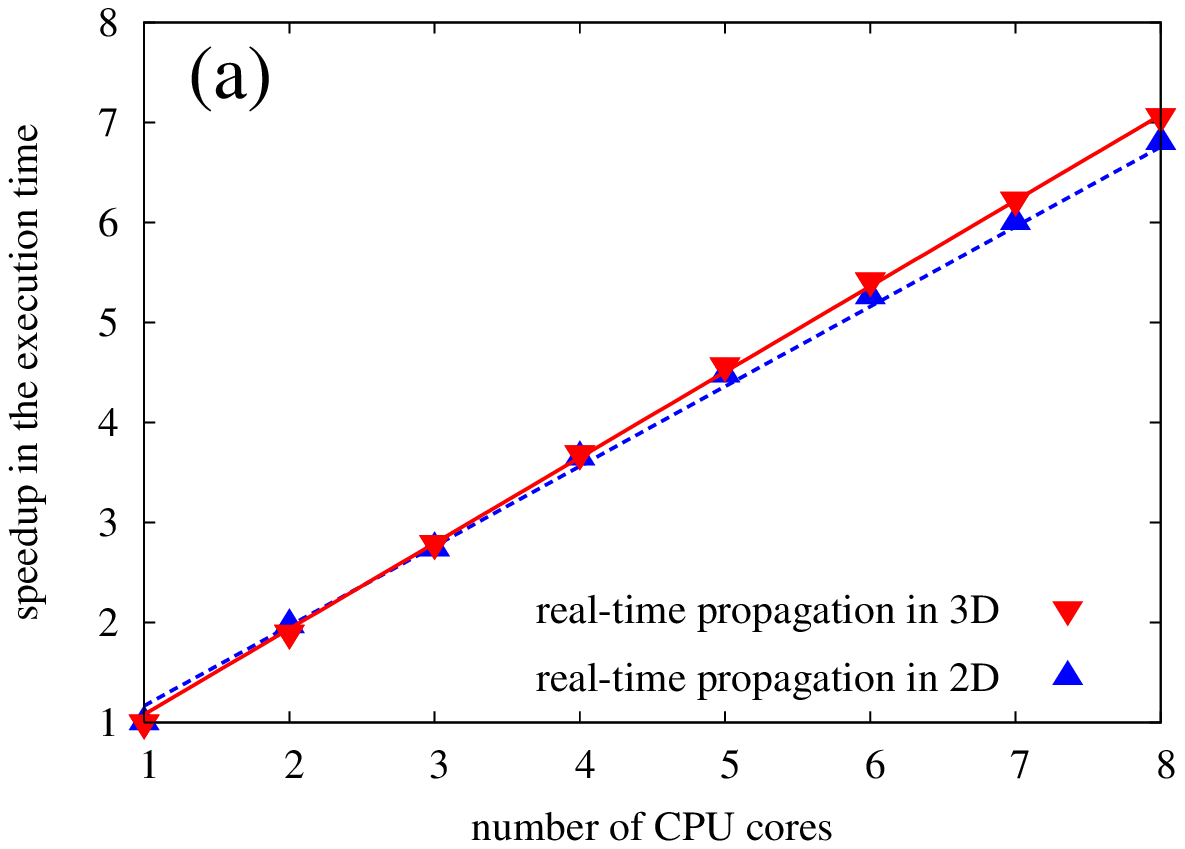}
\includegraphics[width=7cm]{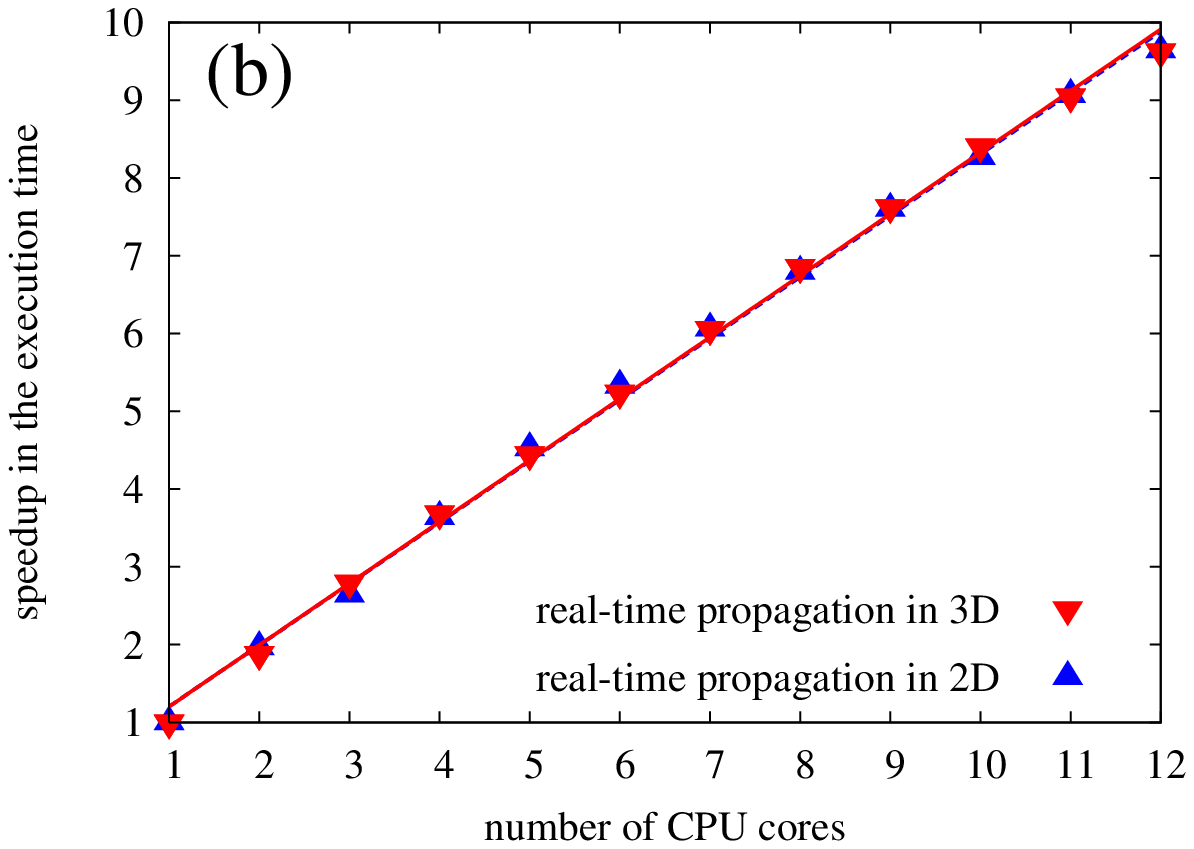}
\caption{(Colour online) Speedup in the execution time of realtime2d-th and realtime3d-th threaded (OpenMP parallelized) programs as a function of the number of CPU cores used. The results are obtained: (a) on an 8-core machine with 2 x quad-core Intel Nehalem E5540 CPU at 2.53~GHz, using the icc compiler,  (b) on a 12-core machine with 2 x  six-core Intel Nehalem X5650 CPU at 2.66~GHz, using the pgcc compiler. The spatial grid sizes used are 2000x2000 (realtime2d-th) and 1000x1000x300 (realtime3d-th).}
\label{fig1}
\end{center}
\end{figure}
\\
{\em Summary of revisions:} All Fortran programs from the previous 
version [1] are translated to C and named in the same way. The structure 
of all programs is identical. We have introduced the use of 
comprehensive input files, where all parameters are explained in detail 
and can be set by a user. We have also included makefiles with tested 
and verified settings for GNU's gcc compiler, Intel's icc compiler, IBM's xlc compiler,
PGI's pgcc compiler, and Oracle's suncc (former Sun's) compiler. In addition to this, 6 new threaded (OpenMP 
parallelized) programs are supplied (imagtime2d-th, imagtime3d-th, 
imagtimeaxial-th, realtime2d-th, realtime3d-th, realtimeaxial-th) for 
algorithms involving two or three space variables. They are written by 
OpenMP-parallelizing the most computationally demanding loops in 
functions performing time evolution (calcnu, calclux, calcluy, calcluz), 
normalization (calcnorm), and calculation of physical quantities 
(calcmuen, calcrms). Since some of the dynamically allocated array 
variables are used within such loops, they had to be made private for 
each thread. This was done by allocating matrices instead of arrays, 
with the first index in all such matrices corresponding to a thread 
number.\\ 
{\em Additional comments:} This package consists of 18 programs, see 
ÒProgram titleÓ above, out of which 12 programs (i, ii, iv, v, vii, ix, 
x, xi, xiii, xiv, xvi, xviii) are serial, while 6 programs (iii, vi, 
viii, xii, xv, xvii) are threaded (OpenMP parallelized). For the 
particular purpose of each program, please see descriptions below.\\
{\em Running time:} All running times given in descriptions below refer to programs compiled with gcc on quad-core Intel Xeon X5460 at 3.16~GHz (CPU1), and programs compiled with icc on quad-core Intel Nehalem E5540 at 2.53~GHz (CPU2). With the supplied input files, running times on CPU1 are: 5 minutes (i, iv, ix, xii, xiii, xvii, xviii), 10 minutes (viii, xvi), 15 minutes (iii, x, xi), 30 minutes (ii, vi, vii), 2 hours (v), 4 hours (xv), 15 hours (xiv). On CPU2, running times are: 5 minutes (i, iii, iv, viii, ix, xii, xiii, xvi, xvii, xviii), 10 minutes (vi, x, xi), 20 minutes (ii, vii), 1 hour (v), 2 hours (xv), 12 hours (xiv).\\

\noindent
{\bf New version program summary (i)}
\\

\noindent
{\em Program title:} imagtime1d\\
{\em Title of electronic files:} imagtime1d.c, imagtime1d.h\\
{\em Computer:} Any modern computer with C language compiler installed\\
{\em Maximum RAM memory:} 4 MByte\\
{\em Programming language used:} C\\
{\em Typical running time:} 2 minutes (CPU1), 1 minute (CPU2)\\
{\em Nature of physical problem:} This program is designed to solve the
time-dependent GP nonlinear partial differential equation
in one space dimension with a harmonic trap. The GP
equation describes the properties of a dilute trapped Bose-Einstein
condensate.\\
{\em Method of solution:} The time-dependent GP equation is 
solved by the split-step Crank-Nicolson method by discretizing in space 
and time. The discretized equation is then solved by propagation in 
imaginary time over small time steps.  The method yields solutions of 
stationary problems.
\\

\noindent
{\bf New version program summary (ii)}
\\

\noindent
{\em Program title:} imagtime2d\\
{\em Title of electronic files:} imagtime2d.c, imagtime2d.h\\
{\em Computer:} Any modern computer with C language compiler installed\\
{\em Maximum RAM memory:} 80 MByte\\
{\em Programming language used:} C\\
{\em Typical running time:} 30 minutes (CPU1), 20 minutes (CPU2)\\
{\em Nature of physical problem:} This program is designed to solve the
time-dependent GP nonlinear partial differential equation
in two space dimensions with an anisotropic trap. The GP
equation describes the properties of a dilute trapped Bose-Einstein
condensate.\\
{\em Method of solution:} The time-dependent GP equation is 
solved by the split-step Crank-Nicolson method by discretizing in space 
and time. The discretized equation is then solved by propagation in 
imaginary time over small time steps.  The method yields solutions of 
stationary problems.  
\\

\noindent
{\bf New version program summary (iii)}
\\

\noindent
{\em Program title:} imagtime2d-th\\
{\em Title of electronic files:} imagtime2d-th.c, imagtime2d-th.h\\
{\em Computer:} Any modern computer with C language compiler installed\\
{\em Maximum RAM memory:} 80 MByte\\
{\em Programming language used:} C/OpenMP\\
{\em Typical running time:} 15 minutes (CPU1), 5 minutes (CPU2)\\
{\em Nature of physical problem:} This program is designed to solve the
time-dependent GP nonlinear partial differential equation
in two space dimensions with an anisotropic trap. The GP
equation describes the properties of a dilute trapped Bose-Einstein
condensate.\\
{\em Method of solution:} The time-dependent GP equation is 
solved by the split-step Crank-Nicolson method by discretizing in space 
and time. The discretized equation is then solved by propagation in 
imaginary time over small time steps.  The method yields solutions of 
stationary problems.  
\\

\noindent
{\bf New version program summary (iv)}
\\

\noindent
{\em Program title:} imagtimecir\\
{\em Title of electronic files:} imagtimecir.c, imagtimecir.h\\
{\em Computer:} Any modern computer with C language compiler installed\\
{\em Maximum RAM memory:} 2 MByte\\
{\em Programming language used:} C\\
{\em Typical running time:} 2 minutes (CPU1), 1 minute (CPU2)\\
{\em Nature of physical problem:} This program is designed to solve the
time-dependent GP nonlinear partial differential equation
in two space dimensions with a circularly-symmetric trap. The
GP equation describes the properties of a dilute trapped
Bose-Einstein condensate.\\
{\em Method of solution:} The time-dependent GP equation is 
solved by the split-step Crank-Nicolson method by discretizing in space 
and time. The discretized equation is then solved by propagation in 
imaginary time over small time steps.  The method yields solutions of 
stationary problems.
\\

\noindent
{\bf New version program summary (v)}
\\

\noindent
{\em Program title:} imagtime3d\\
{\em Title of electronic files:} imagtime3d.c, imagtime3d.h\\
{\em Computer:} Any modern computer with C language compiler installed\\
{\em Maximum RAM memory:} 1.2 GByte\\
{\em Programming language used:} C\\
{\em Typical running time:} 1.5 hours (CPU1), 1 hour (CPU2)\\
{\em Nature of physical problem:} This program is designed to solve the
time-dependent GP nonlinear partial differential equation
in three space dimensions with an anisotropic trap. The GP
equation describes the properties of a dilute trapped Bose-Einstein
condensate.\\
{\em Method of solution:} The time-dependent GP equation is 
solved by the split-step Crank-Nicolson method by discretizing in space 
and time. The discretized equation is then solved by propagation in 
imaginary time over small time steps.  The method yields solutions of 
stationary problems. 
\\

\noindent
{\bf New version program summary (vi)}
\\

\noindent
{\em Program title:} imagtime3d-th\\
{\em Title of electronic files:} imagtime3d-th.c, imagtime3d-th.h\\
{\em Computer:} Any modern computer with C language compiler installed\\
{\em Maximum RAM memory:} 1.2 GByte\\
{\em Programming language used:} C/OpenMP\\
{\em Typical running time:} 25 minutes (CPU1), 10 minutes (CPU2)\\
{\em Nature of physical problem:} This program is designed to solve the
time-dependent GP nonlinear partial differential equation
in three space dimensions with an anisotropic trap. The GP
equation describes the properties of a dilute trapped Bose-Einstein
condensate.\\
{\em Method of solution:} The time-dependent GP equation is 
solved by the split-step Crank-Nicolson method by discretizing in space 
and time. The discretized equation is then solved by propagation in 
imaginary time over small time steps.  The method yields solutions of 
stationary problems. 
\\

\noindent
{\bf New version program summary (vii)}
\\

\noindent
{\em Program title:} imagtimeaxial\\
{\em Title of electronic files:} imagtimeaxial.c, imagtimeaxial.h\\
{\em Computer:} Any modern computer with C language compiler installed\\
{\em Maximum RAM memory:} 32 MByte\\
{\em Programming language used:} C\\
{\em Typical running time:} 30 minutes (CPU1), 20 minutes (CPU2)\\
{\em Nature of physical problem:} This program is designed to solve the
time-dependent GP nonlinear partial differential equation
in three space dimensions with an axially-symmetric trap. The
GP equation describes the properties of a dilute trapped
Bose-Einstein condensate.\\
{\em Method of solution:} The time-dependent GP equation is 
solved by the split-step Crank-Nicolson method by discretizing in space 
and time. The discretized equation is then solved by propagation in 
imaginary time over small time steps.  The method yields solutions of 
stationary problems. 
\\

\noindent
{\bf New version program summary (viii)}
\\

\noindent
{\em Program title:} imagtimeaxial-th\\
{\em Title of electronic files:} imagtimeaxial-th.c, imagtimeaxial-th.h\\
{\em Computer:} Any modern computer with C language compiler installed\\
{\em Maximum RAM memory:} 32 MByte\\
{\em Programming language used:} C/OpenMP\\
{\em Typical running time:} 10 minutes (CPU1), 5 minutes (CPU2)\\
{\em Nature of physical problem:} This program is designed to solve the
time-dependent GP nonlinear partial differential equation
in three space dimensions with an axially-symmetric trap. The
GP equation describes the properties of a dilute trapped
Bose-Einstein condensate.\\
{\em Method of solution:} The time-dependent GP equation is 
solved by the split-step Crank-Nicolson method by discretizing in space 
and time. The discretized equation is then solved by propagation in 
imaginary time over small time steps.  The method yields solutions of 
stationary problems. 
\\

\noindent
{\bf New version program summary (ix)}
\\

\noindent
{\em Program title:} imagtimesph\\
{\em Title of electronic files:} imagtimesph.c, imagtimesph.h\\
{\em Computer:} Any modern computer with C language compiler installed\\
{\em Maximum RAM memory:} 2.5 MByte\\
{\em Programming language used:} C\\
{\em Typical running time:} 2 minutes (CPU1), 1 minute (CPU2)\\
{\em Nature of physical problem:} This program is designed to solve the
time-dependent GP nonlinear partial differential equation
in three space dimensions with a spherically-symmetric trap. The
GP equation describes the properties of a dilute trapped
Bose-Einstein condensate.\\
{\em Method of solution:} The time-dependent GP equation is 
solved by the split-step Crank-Nicolson method by discretizing in space 
and time. The discretized equation is then solved by propagation in 
imaginary time over small time steps.  The method yields solutions of 
stationary problems.
\\

\noindent
{\bf New version program summary (x)}
\\

\noindent
{\em Program title:} realtime1d\\
{\em Title of electronic files:} realtime1d.c, realtime1d.h\\
{\em Computer:} Any modern computer with C language compiler installed\\
{\em Maximum RAM memory:} 4 MByte\\
{\em Programming language used:} C\\
{\em Typical running time:} 15 minutes (CPU1), 10 minutes (CPU2)\\
{\em Nature of physical problem:} This program is designed to solve the
time-dependent GP nonlinear partial differential equation
in one space dimension with a harmonic trap. The GP
equation describes the properties of a dilute trapped Bose-Einstein
condensate.\\
{\em Method of solution:} The time-dependent GP equation is 
solved by the split-step Crank-Nicolson method by discretizing in space 
and time. The discretized equation is then solved by propagation in 
real time over small time steps.  The method yields solutions of 
stationary and non-stationary problems. 
\\

\noindent
{\bf New version program summary (xi)}
\\

\noindent
{\em Program title:} realtime2d\\
{\em Title of electronic files:} realtime2d.c, realtime2d.h\\
{\em Computer:} Any modern computer with C language compiler installed\\
{\em Maximum RAM memory:} 8 MByte\\
{\em Programming language used:} C\\
{\em Typical running time:} 15 minutes (CPU1), 10 minutes (CPU2)\\
{\em Nature of physical problem:} This program is designed to solve the
time-dependent GP nonlinear partial differential equation
in two space dimensions with an anisotropic trap. The GP
equation describes the properties of a dilute trapped Bose-Einstein
condensate.\\
{\em Method of solution:} The time-dependent GP equation is 
solved by the split-step Crank-Nicolson method by discretizing in space 
and time. The discretized equation is then solved by propagation in 
real time over small time steps.  The method yields solutions of 
stationary and non-stationary problems. 
\\

\noindent
{\bf New version program summary (xii)}
\\

\noindent
{\em Program title:} realtime2d-th\\
{\em Title of electronic files:} realtime2d-th.c, realtime2d-th.h\\
{\em Computer:} Any modern computer with C language compiler installed\\
{\em Maximum RAM memory:} 8 MByte\\
{\em Programming language used:} C/OpenMP\\
{\em Typical running time:} 5 minutes (CPU1), 2 minutes (CPU2)\\
{\em Nature of physical problem:} This program is designed to solve the
time-dependent GP nonlinear partial differential equation
in two space dimensions with an anisotropic trap. The GP
equation describes the properties of a dilute trapped Bose-Einstein
condensate.\\
{\em Method of solution:} The time-dependent GP equation is 
solved by the split-step Crank-Nicolson method by discretizing in space 
and time. The discretized equation is then solved by propagation in 
real time over small time steps.  The method yields solutions of 
stationary and non-stationary problems. 
\\

\noindent
{\bf New version program summary (xiii)}
\\

\noindent
{\em Program title:} realtimecir\\
{\em Title of electronic files:} realtimecir.c, realtimecir.h\\
{\em Computer:} Any modern computer with C language compiler installed\\
{\em Maximum RAM memory:} 3 MByte\\
{\em Programming language used:} C\\
{\em Typical running time:} 5 minutes (CPU1), 5 minutes (CPU2)\\
{\em Nature of physical problem:} This program is designed to solve the
time-dependent GP nonlinear partial differential equation
in two space dimensions with a circularly-symmetric trap. The
GP equation describes the properties of a dilute trapped
Bose-Einstein condensate.\\
{\em Method of solution:} The time-dependent GP equation is 
solved by the split-step Crank-Nicolson method by discretizing in space 
and time. The discretized equation is then solved by propagation in 
real time over small time steps.  The method yields solutions of 
stationary and non-stationary problems. 
\\

\noindent
{\bf New version program summary (xiv)}
\\

\noindent
{\em Program title:} realtime3d\\
{\em Title of electronic files:} realtime3d.c, realtime3d.h\\
{\em Computer:} Any modern computer with C language compiler installed\\
{\em Maximum RAM memory:} 700 MByte\\
{\em Programming language used:} C\\
{\em Typical running time:} 15 hours (CPU1), 12 hours (CPU2)\\
{\em Nature of physical problem:} This program is designed to solve the
time-dependent GP nonlinear partial differential equation
in three space dimensions with an anisotropic trap. The GP
equation describes the properties of a dilute trapped Bose-Einstein
condensate.\\
{\em Method of solution:} The time-dependent GP equation is
solved by the split-step Crank-Nicolson method by discretizing in space
and time. The discretized equation is then solved by propagation in real
time over small time steps.  The method yields solutions of
stationary and non-stationary problems.
\\

\noindent
{\bf New version program summary (xv)}
\\

\noindent
{\em Program title:} realtime3d-th\\
{\em Title of electronic files:} realtime3d-th.c, realtime3d-th.h\\
{\em Computer:} Any modern computer with C language compiler installed\\
{\em Maximum RAM memory:} 700 MByte\\
{\em Programming language used:} C/OpenMP\\
{\em Typical running time:} 4 hours (CPU1), 1.8 hours (CPU2)\\
{\em Nature of physical problem:} This program is designed to solve the
time-dependent GP nonlinear partial differential equation
in three space dimensions with an anisotropic trap. The GP
equation describes the properties of a dilute trapped Bose-Einstein
condensate.\\
{\em Method of solution:} The time-dependent GP equation is
solved by the split-step Crank-Nicolson method by discretizing in space
and time. The discretized equation is then solved by propagation in real
time over small time steps.  The method yields solutions of
stationary and non-stationary problems.
\\

\noindent
{\bf New version program summary (xvi)}
\\

\noindent
{\em Program title:} realtimeaxial\\
{\em Title of electronic files:} realtimeaxial.c, realtimeaxial.h\\
{\em Computer:} Any modern computer with C language compiler installed\\
{\em Maximum RAM memory:} 4 MByte\\
{\em Programming language used:} C\\
{\em Typical running time:} 10 minutes (CPU1), 5 minutes (CPU2)\\
{\em Nature of physical problem:} This program is designed to solve the
time-dependent GP nonlinear partial differential equation
in three space dimensions with an axially-symmetric trap. The
GP equation describes the properties of a dilute trapped
Bose-Einstein condensate.\\
{\em Method of solution:} The time-dependent GP equation is 
solved by the split-step Crank-Nicolson method by discretizing in space 
and time. The discretized equation is then solved by propagation in 
real time over small time steps.  The method yields solutions of 
stationary and non-stationary problems. 
\\

\noindent
{\bf New version program summary (xvii)}
\\

\noindent
{\em Program title:} realtimeaxial-th\\
{\em Title of electronic files:} realtimeaxial-th.c, realtimeaxial-th.h\\
{\em Computer:} Any modern computer with C language compiler installed\\
{\em Maximum RAM memory:} 4 MByte\\
{\em Programming language used:} C/OpenMP\\
{\em Typical running time:} 5 minutes (CPU1), 1 minute (CPU2)\\
{\em Nature of physical problem:} This program is designed to solve the
time-dependent GP nonlinear partial differential equation
in three space dimensions with an axially-symmetric trap. The
GP equation describes the properties of a dilute trapped
Bose-Einstein condensate.\\
{\em Method of solution:} The time-dependent GP equation is 
solved by the split-step Crank-Nicolson method by discretizing in space 
and time. The discretized equation is then solved by propagation in 
real time over small time steps.  The method yields solutions of 
stationary and non-stationary problems. 
\\

\noindent
{\bf New version program summary (xviii)}
\\

\noindent
{\em Program title:} realtimesph\\
{\em Title of electronic files:} realtimesph.c, realtimesph.h\\
{\em Computer:} Any modern computer with C language compiler installed\\
{\em Maximum RAM memory:} 2.5 MByte\\
{\em Programming language used:} C\\
{\em Typical running time:} 5 minutes (CPU1), 5 minutes (CPU2)\\
{\em Nature of physical problem:} This program is designed to solve the
time-dependent GP nonlinear partial differential equation
in three space dimensions with a spherically-symmetric trap. The
GP equation describes the properties of a dilute trapped
Bose-Einstein condensate.\\
{\em Method of solution:} The time-dependent GP equation is 
solved by the split-step Crank-Nicolson method by discretizing in space 
and time. The discretized equation is then solved by propagation in 
real time over small time steps.  The method yields solutions of 
stationary and non-stationary problems. 
\\

\end{small}

\begin{thebibliography}{19}

\bibitem{1}
P. Muruganandam, S.~K. Adhikari, Fortran programs for the time-dependent Gross-Pitaevskii equation in a fully anisotropic trap,
Comput. Phys. Commun. 180 (2009) 1888.

\bibitem{2}
G. Mazzarella, L. Salasnich, Collapse of triaxial bright solitons in atomic Bose-Einstein condensates, 
Phys. Lett. A 373 (2009) 4434.

\bibitem{3}
Y. Cheng, S.~K. Adhikari, Symmetry breaking in a localized interacting binary Bose-Einstein condensate in a bichromatic optical lattice, 
Phys. Rev. A 81 (2010) 023620;

S.~K. Adhikari, H. Lu, H. Pu, Self-trapping of a Fermi superfluid in a double-well potential in the Bose-Einstein-condensate-unitarity crossover, 
Phys. Rev. A 80 (2009)  063607.

\bibitem{4}
S. Gautam, D. Angom, Rayleigh-Taylor instability in binary condensates, 
Phys. Rev. A 81 (2010) 053616;

S. Gautam, D. Angom,
Ground state geometry of binary condensates in axisymmetric traps, J. Phys. B 43 (2010) 095302;

S. Gautam, P.  Muruganandam, D. Angom, Position swapping and pinching in Bose-Fermi mixtures with two-color optical Feshbach resonances,
Phys. Rev. A  83  (2011) 023605. 

\bibitem{5}
S.~K. Adhikari, B.~A. Malomed, L. Salasnich,  F. Toigo, 
Spontaneous symmetry breaking of Bose-Fermi mixtures in double-well potentials, 
Phys. Rev. A 81 (2010) 053630.

\bibitem{6}
G.~K. Chaudhary,  R. Ramakumar, Collapse dynamics of a (176)Yb-(174)Yb Bose-Einstein condensate, Phys. Rev. A 81 (2010) 063603.

\bibitem{7}
S. Sabari, R.~V.~J. Raja, K. Porsezian,  P. Muruganandam, Stability of trapless Bose-Einstein condensates with two- and three-body interactions,
 J. Phys. B 43 (2010) 125302.

\bibitem{8}
L.~E. Young-S, L. Salasnich,  S.~K. Adhikari, Dimensional reduction of a binary Bose-Einstein condensate in mixed dimensions, 
Phys. Rev. A 82 (2010) 053601.

\bibitem{9}
 I. Vidanovi\'{c}, A. Bala\v{z}, H. Al-Jibbouri, A. Pelster,
Nonlinear Bose-Einstein-condensate dynamics induced by a harmonic modulation of the s-wave scattering length, 
 Phys. Rev. A 84 (2011) 013618.
 
\bibitem{10}R.~R. Sakhel, A.~R.  Sakhel, H.~B.  Ghassib, Self-interfering matter-wave patterns generated by a moving laser obstacle in a two-dimensional Bose-Einstein condensate inside a power trap cut off by box potential boundaries, Phys. Rev. A 84 (2011) 033634.
 

\bibitem{11}
A. Bala\v{z},  A.~I. Nicolin, Faraday waves in binary nonmiscible Bose-Einstein condensates, Phys. Rev. A 85 (2012) 023613;

A.~I. Nicolin, Variational treatment of Faraday waves in inhomogeneous BoseÐEinstein condensates, Physica A 391 (2012) 1062;

A.~I. Nicolin, Resonant wave formation in Bose-Einstein condensates, Phys. Rev. E 84 (2011) 056202; 

A.~I. Nicolin, Faraday waves in  Bose-Einstein condensates subject to anisotropic transverse confinement, Rom. Rep. Phys. 63 (2011) 1329;

A.~I. Nicolin, M.~C. Raportaru, Faraday waves in high-density cigar-shaped BoseÐEinstein condensates, Physica A 389 (2010) 4663.

\bibitem{12}
S. Yang, M. Al-Amri, J. Evers, M.~S. Zubairy, 
Controllable optical switch using a Bose-Einstein condensate in an optical cavity, 
Phys. Rev. A 83 (2011) 053821.

\bibitem{13}
W. Hua, X.-S. Liu, Dynamics of cubic and quintic nonlinear Schrodinger equations, Acta Phys. Sin. 60 (2011) 110210.

\bibitem{14}
Z. Sun,  W. Yang, An exact short-time solver for the time-dependent Schrodinger equation, 
J. Chem. Phys. 134 (2011) 041101.

\bibitem{15}
A. Bala\v{z}, I. Vidanovi\'{c}, A. Bogojevi\'{c}, A. Beli\'{c},  A. Pelster, 
Fast converging path integrals for time-dependent potentials: I. Recursive calculation of short-time expansion of the propagator, 
J. Stat. Mech.-Theory Exp. (2011) P03004.

\bibitem{16}
W.~B. Cardoso, A.~T. Avelar, D. Bazeia,
One-dimensional reduction of the three-dimenstional Gross-Pitaevskii equation with two- and three-body interactions,  Phys. Rev. E 83 (2011) 036604.
 

\end{thebibliography}
\end{document}